\documentstyle[twoside]{article}
\catcode`\@=11
\long\def\@makefntext#1{
\protect\noindent \hbox to 3.2pt {\hskip-.9pt  

$^{{\eightrm\@thefnmark}}$\hfil}#1\hfill}       

\def\thefootnote{\fnsymbol{footnote}}
\def\@makefnmark{\hbox to 0pt{$^{\@thefnmark}$\hss}}    

\def\ps@myheadings{\let\@mkboth\@gobbletwo
\def\@oddhead{\hbox{}
\rightmark\hfil\eightrm\thepage}   

\def\@oddfoot{}\def\@evenhead{\eightrm\thepage\hfil
\leftmark\hbox{}}\def\@evenfoot{}
\def\sectionmark##1{}\def\subsectionmark##1{}}

\oddsidemargin=\evensidemargin
\renewcommand{\thefootnote}{\fnsymbol{footnote}}

\newcounter{sectionc}\newcounter{subsectionc}\newcounter{subsubsectionc}
\renewcommand{\section}[1] {\vspace{12pt}\addtocounter{sectionc}{1} 

\setcounter{subsectionc}{0}\setcounter{subsubsectionc}{0}\noindent 

    {\tenbf\thesectionc. #1}\par\vspace{5pt}}
\renewcommand{\subsection}[1] {\vspace{12pt}\addtocounter{subsectionc}{1} 

    \setcounter{subsubsectionc}{0}\noindent 

    {\bf\thesectionc.\thesubsectionc. {\kern1pt \bfit #1}}\par\vspace{5pt}}
\renewcommand{\subsubsection}[1] {\vspace{12pt}\addtocounter{subsubsectionc}{1}
    \noindent{\tenrm\thesectionc.\thesubsectionc.\thesubsubsectionc.
    {\kern1pt \tenit #1}}\par\vspace{5pt}}
\newcommand{\nonumsection}[1] {\vspace{12pt}\noindent{\tenbf #1}
    \par\vspace{5pt}}

\newcounter{appendixc}
\newcounter{subappendixc}[appendixc]
\newcounter{subsubappendixc}[subappendixc]
\renewcommand{\thesubappendixc}{\Alph{appendixc}.\arabic{subappendixc}}
\renewcommand{\thesubsubappendixc}
    {\Alph{appendixc}.\arabic{subappendixc}.\arabic{subsubappendixc}}

\renewcommand{\appendix}[1] {\vspace{12pt}
        \refstepcounter{appendixc}
        \setcounter{figure}{0}
        \setcounter{table}{0}
        \setcounter{lemma}{0}
        \setcounter{theorem}{0}
        \setcounter{corollary}{0}
        \setcounter{definition}{0}
        \setcounter{equation}{0}
        \renewcommand{\thefigure}{\Alph{appendixc}.\arabic{figure}}
        \renewcommand{\thetable}{\Alph{appendixc}.\arabic{table}}
        \renewcommand{\theappendixc}{\Alph{appendixc}}
        \renewcommand{\thelemma}{\Alph{appendixc}.\arabic{lemma}}
        \renewcommand{\thetheorem}{\Alph{appendixc}.\arabic{theorem}}
        \renewcommand{\thedefinition}{\Alph{appendixc}.\arabic{definition}}
        \renewcommand{\thecorollary}{\Alph{appendixc}.\arabic{corollary}}
        \renewcommand{\theequation}{\Alph{appendixc}.\arabic{equation}}
        \noindent{\tenbf Appendix \theappendixc #1}\par\vspace{5pt}}
\newcommand{\subappendix}[1] {\vspace{12pt}
        \refstepcounter{subappendixc}
        \noindent{\bf Appendix \thesubappendixc. {\kern1pt \bfit #1}}
    \par\vspace{5pt}}
\newcommand{\subsubappendix}[1] {\vspace{12pt}
        \refstepcounter{subsubappendixc}
        \noindent{\rm Appendix \thesubsubappendixc. {\kern1pt \tenit #1}}
    \par\vspace{5pt}}

\newcommand{\textlineskip}{\baselineskip=13pt}
\newcommand{\smalllineskip}{\baselineskip=10pt}

\def\eightcirc{
\begin{picture}(0,0)
\put(4.4,1.8){\circle{6.5}}
\end{picture}}
\def\eightcopyright{\eightcirc\kern2.7pt\hbox{\eightrm c}}

\def\abstracts#1#2#3{{
    \centering{\begin{minipage}{4.5in}\baselineskip=10pt\footnotesize
    \parindent=0pt #1\par 

    \parindent=15pt #2\par
    \parindent=15pt #3
    \end{minipage}}\par}}

\renewenvironment{thebibliography}[1]
    {\frenchspacing
     \ninerm\baselineskip=11pt
     \begin{list}{\arabic{enumi}.}
    {\usecounter{enumi}\setlength{\parsep}{0pt}
     \setlength{\leftmargin 12.7pt}{\rightmargin 0pt} 
     \setlength{\itemsep}{0pt} \settowidth
    {\labelwidth}{#1.}\sloppy}}{\end{list}}

\newcounter{itemlistc}
\newcounter{romanlistc}
\newcounter{alphlistc}
\newcounter{arabiclistc}

\newcommand{\fcaption}[1]{
        \refstepcounter{figure}
        \setbox\@tempboxa = \hbox{\footnotesize Fig.~\thefigure. #1}
        \ifdim \wd\@tempboxa > 5in
           {\begin{center}
        \parbox{5in}{\footnotesize\smalllineskip Fig.~\thefigure. #1}
            \end{center}}
        \else
             {\begin{center}
             {\footnotesize Fig.~\thefigure. #1}
              \end{center}}
        \fi}

\newcommand{\tcaption}[1]{
        \refstepcounter{table}
        \setbox\@tempboxa = \hbox{\footnotesize Table~\thetable. #1}
        \ifdim \wd\@tempboxa > 5in
           {\begin{center}
        \parbox{5in}{\footnotesize\smalllineskip Table~\thetable. #1}
            \end{center}}
        \else
             {\begin{center}
             {\footnotesize Table~\thetable. #1}
              \end{center}}
        \fi}

\def\@citex[#1]#2{\if@filesw\immediate\write\@auxout
    {\string\citation{#2}}\fi
\def\@citea{}\@cite{\@for\@citeb:=#2\do
    {\@citea\def\@citea{,}\@ifundefined
    {b@\@citeb}{{\bf ?}\@warning
    {Citation `\@citeb' on page \thepage \space undefined}}
    {\csname b@\@citeb\endcsname}}}{#1}}

\newif\if@cghi
\def\cite{\@cghitrue\@ifnextchar [{\@tempswatrue
    \@citex}{\@tempswafalse\@citex[]}}
\def\citelow{\@cghifalse\@ifnextchar [{\@tempswatrue
    \@citex}{\@tempswafalse\@citex[]}}
\def\@cite#1#2{{$\null^{#1}$\if@tempswa\typeout
    {IJCGA warning: optional citation argument 

    ignored: `#2'} \fi}}

\def\pmb#1{\setbox0=\hbox{#1}
    \kern-.025em\copy0\kern-\wd0
    \kern.05em\copy0\kern-\wd0
    \kern-.025em\raise.0433em\box0}

\def\fnt#1#2{\footnotetext{\kern-.3em
    {$^{\mbox{\scriptsize #1}}$}{#2}}}

\def\fpage#1{\begingroup
\voffset=.3in
\thispagestyle{empty}\begin{table}[b]\centerline{\footnotesize #1}
    \end{table}\endgroup}

\def\runninghead#1#2{\pagestyle{myheadings}
\markboth{{\protect\footnotesize\it{\quad #1}}\hfill}
{\hfill{\protect\footnotesize\it{#2\quad}}}}
\font\tenrm=cmr10
\font\tenit=cmti10 

\font\tenbf=cmbx10
\font\bfit=cmbxti10 at 10pt
\font\ninerm=cmr9

\font\eightrm=cmr8

\def\qed{\hbox{${\vcenter{\vbox{            
   \hrule height 0.4pt\hbox{\vrule width 0.4pt height 6pt
   \kern5pt\vrule width 0.4pt}\hrule height 0.4pt}}}$}}

\renewcommand{\thefootnote}{\fnsymbol{footnote}}    

\begin{document}
\runninghead{Integrable Models, SUSY Gauge Theories, and String Theory}  
{Integrable Models, SUSY Gauge Theories, and String Theory}

\normalsize\textlineskip
\thispagestyle{empty}
\setcounter{page}{1}


\vspace*{0.88truein}

\fpage{1}
\centerline{\bf  INTEGRABLE MODELS,  SUSY GAUGE THEORIES}
\vspace*{0.035truein}
\centerline{\bf AND STRING THEORY\footnote
{Talk given at the {\it Workshop on Low Dimensional Field Theory},
Aug. 5-17, 1996, Telluride, CO, USA.}}
\vspace*{0.37truein}
\centerline{\footnotesize SOONKEON NAM\footnote{
nam@nms.kyunghee.ac.kr}}
\vspace*{0.015truein}
\centerline{\footnotesize\it Department of Physics and
Research Institute for Basic Sciences}
\baselineskip=10pt
\centerline{\footnotesize\it Kyung Hee University, Seoul,130-701 ,
KOREA}
\vspace*{0.225truein}

\vspace*{0.21truein}
\abstracts{
We consider the close relation between duality in $N=2$ SUSY gauge theories
and integrable models. Various integrable models ranging from Toda lattices, Calogero models,
spinning tops, and  spin chains are related to  the quantum moduli space of  vacua of $N=2$ 
SUSY gauge  theories. In  particular, $SU(3)$ gauge  theories with two  flavors of  massless 
quarks
in  the   fundamental  representation   can  be   related  to   the  spectral  curve   of  the 
Goryachev-Chaplygin top, which is a Nahm's equation in disguise. 
This can be generalized  to the cases with {\it massive} quarks,  and $N_f = 0,1,2$, where a 
system with seven dimensional phase space has the  relevant hyperelliptic curve appear in the 
Painlev\'e test.   To understand  the stringy origin of  the integrability of  these theories we 
obtain  exact  nonperturbative  point  particle  limit  of  type  II  string  compactified on  a 
Calabi-Yau manifold,  which gives  the hyperelliptic curve  of $SU(2)$ QCD  with $N_f  =1$ 
hypermultiplet.}{}{}


\vspace*{1pt}\textlineskip  
\section{Introduction}  
\vspace*{-0.5pt}
\noindent

\textheight=7.8truein
\setcounter{footnote}{0}
\renewcommand{\thefootnote}{\alph{footnote}}

\def\beq{\begin{equation}}
\def\eeq{\end{equation}}
\def\bea{\begin{eqnarray}}
\def\eea{\end{eqnarray}}
\renewcommand{\arraystretch}{1.5}
\def\ba{\begin{array}}
\def\ea{\end{array}}
\def\bce{\begin{center}}
\def\ece{\end{center}}
\def\nn{\noindent}
\def\nonu{\nonumber}
\def\pbx{\partial_x}

\def\ptl{\partial}
\def\al{\alpha}
\def\be{\beta}
\def\ga{\gamma} 
\def\Ga{\Gamma}
\def\de{\delta} \def\De{\Delta}
\def\ep{\epsilon}
\def\vep{\varepsilon}
\def\ze{\zeta}
\def\et{\eta}
\def\th{\theta} \def\Th{\Theta}
\def\vth{\vartheta}
\def\io{\iota}
\def\ka{\kappa}
\def\la{\lambda} 
\def\La{\Lambda}
\def\rh{\rho}
\def\si{\sigma} \def\Si{\Sigma}
\def\ta{\tau}
\def\up{\upsilon} 
\def\Up{\Upsilon}
\def\ph{\phi} 
\def\Ph{\Phi}
\def\vph{\varphi}
\def\ch{\chi}
\def\ps{\psi} 
\def\Ps{\Psi}
\def\om{\omega} 
\def\Om{\Omega}

\def\lbr{\left(}
\def\rbr{\right)}
\def\half{\frac{1}{2}}
\def\CVO#1#2#3{\!\left( \matrix{ #1 \cr #2 \ #3 \cr} \right)\!}

\def\vol#1{{#1}}
\def\nupha#1{{\it Nucl. Phys. }\vol{#1} }
\def\phlta#1{{\it Phys. Lett. }\vol{#1} }
\def\phyrv#1{{\it Phys. Rev. }\vol{#1} }
\def\PRL#1{{\it Phys. Rev. Lett. } \vol{#1} }
\def\prs#1{{\it Proc. Roc. Soc. }\vol{#1} }
\def\PTP#1{{\it Prog. Theo. Phys. }\vol{#1} }
\def\SJNP#1{{\it Sov. J. Nucl. Phys. }\vol{#1} }
\def\TMP#1{{\it Theor. Math. Phys. }\vol{#1} }
\def\ANNPHY#1{{\it Annals of Phys. }\vol{#1} }
\def\PNAS#1{{\it Proc. Natl. Acad. Sci. USA }\vol{#1} }
\def\CMP#1{{\it Comm. Math. Phys. }\vol{#1} }
\def\MPL#1{{\it Mod. Phys. Lett. }\vol{#1} }

One of the challenging problems in theoretical physics is to understand
nonperturbative behavior of field theories and string theories.
Last several years we have witnessed a very important progress in 
understanding duality of $N=2$ SUSY gauge 
theories\cite{SW}.   For  the first  time, we  have tools  to deal  with exact  nonperturbative 
calculations.  The low  energy  description of  these  theories can  be  encoded by  Riemann 
surfaces and  the integrals of meromorphic one differentials over the periods of them.
Exact effective actions of these theories can be described by holomorphic functions,
so-called prepotentials. With these we can probe the strong coupling limits
of the theories. It would desirable to have a better understanding of these structures, through
other physical systems where similar structures appear.
In fact in the study of  the integrable models in two dimensions,
which were long studied in the hopes of giving insights into higher dimensional systems,
this structure on  Riemann surfaces plays a crucial role\cite{DKN}. 
Among the methods of solving integrable models, in
inverse scattering method we obtain the solitons solutions as potentials
of a quantum mechanics problem, given the scattering data.
The spectral parameter plays the role of the energy.
If we consider the periodic soliton solutions, then the spectral parameter
develops forbidden zones, just as we are familar in solid state physics.
Analytic continuation of the spectral parameter with the forbidden zones
 gives us the Riemann surface with genus $g>0$.
By now there are many works which connect these low-energy effective
theories with known integrable systems.
To relate effective quantum field theories with integrable systems, 
one needs averaging over fast oscillations, i.e. Whitham averaging.
It was analyzed that the periods of the modulated Whitham solution
of periodic Toda lattice give rise to the mass
spectrum in the BPS saturated states\cite{GKMMM,NT}.
For the case of $SU(N_{c})$ gauge theory with a single
hypermultiplet in the {\it adjoint} representation, 
the corresponding integrable system was found
and recognized to be the elliptic spin Calogero model\cite{MW,DW,Mar}, where
short range interaction of Toda lattice is generalized to a long ranged 
integrable interaction.  This connection was developed by identifying the 
coupling constant of Calogero system with the mass of a 
hypermultiplet in the adjoint representation, 
starting from the Lax operator for the Calogero model and
calculating the spectral curve explicitly\cite{IM}.
The integrable system related to 
gauge theories with to {\it massive} hypermultiplets 
in the {\it fundamental} representation was also discussed\cite{AN,Brz,RUS}. 
Here the relevant integrable models are spinning tops and/or spin chain models!
These seemingly different systems share a common mathematical structure.
More  recently,  Seiberg  and  Witten  further  studied  the  $N=4$  gauge  theories   in  3 
dimensions\cite{SW3}
and structures such as Nahm's equations\cite{Nahm},
which appears in the study of moduli space of muiltimonopoles. 
Hyperk\"ahler structures such  as Atiyah-Hitchin space, its  double covering, and Taub-NUT 
spaces with dihedral quotients appear\cite{SW3}.
We can easily be puzzled by the plethora of models all claiming a relation to
the nonperturbative SUSY dynamics.
It is likely that there is a underlying unified point of view.
In fact there is a mapping of monopoles spectral curves from the Nahm's equations
and that of Toda lattices\cite{Sut}.
Furthermore we can map the Nahm's equation to the generalized Kowalevski top\cite{Manas}
extending the bridges among the models.
We can actually extend this and relate the Nahm's equation to the GC top.
Underlying scheme might as well be the self-dual Yang Mills equations:
a wide class of low dimensional integrable models can be obtained from the
self-dual Yang Mills equation\cite{Ward,Ablo}.
The relation between the spin  chain models\cite{RUS} and the spinning tops\cite{AN} is less 
clear,  even though both appear in SW models with massive hypermultiplets.
This clearly invokes a further study.

Motivated by the works in SUSY gauge theories, the duality
really blossomed in the context of string theories\cite{JP}.
Even though the current string duality is very useful in understanding
{\it strong coupling} regime utilizing the weak coupling expansion in the 
dual model, it is at the moment not clear how the intermediate coupling regime will be
described.
Certain self duality might be useful there.
Among these, the $N=2$ type II/heterotic duality in four dimensions 
has been proposed\cite{KV} and further studied in many subsequent
papers.  In fact, it was extended to the $F$-theory/heterotic duality\cite{MV}
in eight dimensions
where the heterotic strings compactified on $T^2$ is dual to 
$F$-theory compactified on $K3$ which admits an elliptic fibration.
Further compactification in six dimensions leads to
the duality between $F$-theories compactified on Calabi Yau(CY) manifolds and
heterotic strings on $K3$.
Among the many ways to check the consistency on this duality 
one can consider the point like limit of four dimensional
$N=2$ SUSY compactifications of heterotic strings, and
see the resulting gauge theory\cite{KKLMV}, which reproduces the
exact field theory results\cite{SW}.
Additional question would be whether one can get also matter
from the point like limit of the string theory compactification.
We would like to see how the $N=2$ SUSY QCD is embedded in this
compactification of string theory\cite{KHU}.

\section{SUSY Gauge Theories}
\noindent

We first consider the $N =2$ SUSY $SU(N_c)$ gauge
theories with $N_c$ colors and $N_f$ flavors. The field content of the 
theories
consists, in terms of $N=1$ superfields, a vector multiplet $W_\al$,
a chiral multiplet $\Ph$, and two chiral superfields $Q^i_a$ and
$\tilde{Q}_{ia}$ where $i= 1, \cdots, N_f$ and $a =1, \cdots, N_c$.
The curve representing the moduli space with
$N_{f} < N_{c}$ case is as follows\cite{HO}:
\bea
y^2=(x^{N_{c}}-\sum_{i=2}^{N_{c}} u_{i} x^{N_{c}-i})^2-
\Lambda_{N_{f}}^{2N_{c}-N_{f}} \prod_{i=1}^{N_{f}} (x+m_{i}),
\label{eq:curve}
\eea
where the moduli $u_{i}$'s are the vacuum expectation 
values of a scalar field
of the $N=2$ chiral multiplet, and $m_i$'s are the bare quark masses. 
It turns out that from the point of view of integrable theory, 
$u_{i}$'s correspond to the integrals of motion. 
The second term in Eq.(\ref{eq:curve}) is due to the instanton corrections.
For the $N_c \leq N_f < 2 N_c$ case, 
the correction due to matter is different\cite{HO}.
By inspection we see that the case of $N_f = 0$ corresponds to the periodic 
Toda lattice with $N_c$-particles,
after an appropriate rescaling of the variables\cite{GKMMM}.
In general the following type of hyperelliptic curve appears 
\beq
y^2 = P_{N_c} (x)^2 - Q_m (x),
\eeq
where $P_n(x)$ and $Q_m (x)$ are polynomials of order $n$ and $m$.
It is natural to ask which integrable theories have such spectral curves.
The form is indicative of Riemann surfaces with punctures as well as genus.
We will start with the known case of $y^2 = P_3 (x)^2 - a x^2$ ($a$ is a constant) 
which corresponds to the so called 
Goryachev-Chaplygin  (GC)   top\footnote{It  was  noted   in  that   there  exists  such   a 
connection\cite{Marsha,AN}.}.

\section{Integrable Models}
\noindent

Let us now review the classical mechanics of rotation of a heavy rigid body 
around a fixed point, which is described by the following Hamiltonian:
\bea
H(M,p)=\frac{M_{1}^2}{2I_{1}} +\frac{M_{2}^2}{2I_{2}}+
\frac{M_{3}^2}{2I_{3}}+\gamma_{1} p_{1}+\gamma_{2} p_{2}+
\gamma_{3} p_{3}.
\label{eq:hamil}
\eea
The phase space of this system is six dimensional: 
$M_i$'s are the components of the angular momentum and $p_i$'s are
the linear momenta.
$I_{i}$'s are the principal
moments of inertia of the body and $\gamma_{i}$'s
are the coordinates of the center of mass. 
There are four known integrable cases for the Hamiltonian in 
Eq.(\ref{eq:hamil}). 
In all these cases there is always one obvious integral of motion, the energy. 
It is necessary to get one extra integral independent of 
the energy for complete integrability 
according to Liouville's theorem\cite{DKN}.

Apart from the better known cases of Euler's and Lagrange's tops,
we have following two other cases:
i) Kowalewski's case: ($ I_{1}=I_{2}=2I_{3}, \gamma_{3}=0$.)
The extra integral can be found by the Painlev\'{e} test or the
Kowalewski's asymptotic
method. Here the symmetry group is $SO(3,2)$.       
ii) Goryachev-Chaplygin's case: 
($I_{1}=I_{2}=4I_{3}, \gamma_{3}=0$.) We need 
$M_{1} p_{1}+M_{2} p_{2}+M_{3} p_{3}=0$, which leads to a new integral of 
motion
together with  $H$ the Hamiltonian and $G$ the GC integral\cite{Koz}.
The Lax operator for the GC top  is given as follows\cite{BK}
where we have written it down in a form useful when comparing to Nahm's equation: 
\bea
&&z L(z)=
\left( \begin{array}{ccc}
0 & -i p_3 & 0 \\
i p_{3} & 0 & p_{2}-i p_{1}  \\
0 & p_{2}+ i p_{1} & 0  \\
\end{array} \right)
\nonu \\
&&\ \ \ \ \ \ \ 
-2iz
\left( \begin{array}{ccc}
0 & 0 & \frac{iM_2 +M_{1}}{2} \\
0 & -M_{3} & 0 \\
\frac{-iM_{2}+ M_{1} }{2}& 0 & M_{3}  \\
\end{array} \right)
+z^2 
\left( \begin{array}{ccc}
0 & 0 & 0\\
0 & 0  & -2 i   \\
0  & 2 i  & 0 \\
\end{array} \right).
\eea
This Lax operator depends on the phase space variables, $M_i, \  p_i$ 
and on the spectral parameter $z$.
Now it is easy to calculate the spectral curve from the equation
${\rm Det}( L(z)-x I ) =0$, which gives the spectral curve
as follows:
\bea
x^3+2 x H -2 i G-x (4 z^2+\frac{\lambda^2}{z^2}) =0,
\eea
where $H=\frac{1}{2} ( M_{1}^2+M_{2}^2+4 M_{3}^2 )-2 p_{1}$ is the
Hamiltonian, and $G=M_{3} (M_{1}^2+M_{2}^2)+2 M_{1} p_{3}$ is the 
GC integral. 
We also have the following constraints: 
\beq
p_{1}^2+p_{2}^2+p_{3}^2=\lambda^2,\;\;\; {\rm and}\;\;\; M_{1} p_{1}+
M_{2} p_{2}+M_{3} p_{3}=0 . 
\eeq
Now we see that the spectral curve
depends on special combinations of $M_i, p_i$'s, which are nothing but 
the integrals of motion.
By introducing variable $y=x (4 z^2-\frac{\lambda^2}{z^2})$, we thus get
\bea
y^2=(x^3+2 H x-2 i G)^2-16 \lambda^2 x^2,
\label{eq:curve1}
\eea
which are the same as the curve for GC top with some rescalings.
To relate this to the curve of SUSY gauge
theory we make the following simple substitutions:
\beq
H \rightarrow -\frac{1}{2} u_{2}, \;\;\;\;\; 
G \rightarrow -\frac{i}{2} u_{3},
\;\;\;\;\; \lambda^2 \rightarrow \frac{1}{16} \Lambda_{2}^{4}.
\eeq
It is easy to see that Eq.(\ref{eq:curve1}) exactly coincides 
with Eq.(\ref{eq:curve}) for the particular
case of $N_{c}=3, N_{f}=2$ and $m_{1}=m_{2}=0$!

As mentioned in the introduction, the GC top can be viewed as the Nahm's equation.
To see this clearly, let us cast the Lax operator of  GC top in a form useful to compare with 
the
Nahm's equation.
We see that we can actually utilize Hitchin's parametrization\cite{Hitchin}
and put the Lax operator as 
follows:
\bea
z L(z) &=& (T_1 + iT_2) - 2iz T_3 +z^2 (T_1 -i T_2),\nonu \\
A(z) &=& T_0 - iT_1 + z(T_1 -i T_2),
\eea
where $T_0$ and $T_i$'s $(i=1,2,3)$ are certain $3\times 3 $ matrices.
Then clearly the Lax equation
$dL/dt = [L,A]$ can be recast as the Nahm's equation:
\beq
\frac{dT_i}{dt} = [T_i, T_0] + \frac{1}{2} \ep_{ijk} [T_j, T_k].
\eeq
The components of $T_i$'s can easily be written down.
Ward observed that the Nahm's equation can be written as a Lax equation, and that
it also can be regarded as a Yang-Baxter equation\cite{Ward2}.
We can gauge away $T_0$.
The Nahm's  equation and can be  mapped into Toda lattice  when $T_i$'s are of  a special 
form\cite{Ward2} 
and also to the generalized Kowalewski's top\cite{Manas}.
The spectral curve of the Nahm's equation obtained from
\beq
{\rm Det}\left(x + i\sum^{3}_{j=1}\eta_j T_j\right) =0,
\eeq
where $\eta_1 = -i(1+z^2), \ \  \eta_2 = 1-z^2, \ \  
\eta_3 = -2z$,
describes the moduli space of  multimonopole configuration, and arises in 
the twister formulation for monopoles\cite{ADHM}.
We note that in a recent work of Seiberg and Witten\cite{SW3} of   SUSY gauge theories on 
compactified three dimensional spacetime, Dancer's spectral curve\cite{Dancer} 
appeared in the context of $SU(2)$  $N_f=1$ case.  

\section{The Massive Case}
\noindent

Since we have seen the intimate relation between the GC top and 
the SUSY $SU(3)$ gauge theory with two flavor
massless hypermultiplets, it is natural for us to extend
this to the massive case. For this purpose we need an integrable system
which has both the GC top and the three body 
Toda lattice as particular limits, because the latter corresponds to
pure gauge theory with no matter.
The Hamiltonian system which realizes this is hard to imagine,
but there exists a system of coupled seven nonlinear differential
equations in mathematical literature\cite{BvM}.
This system has the following nonlinear ``equations of motion":
\bea
&&\dot{z_{1}}  =  -8 z_{7}, \ \ \ \ 
\dot{z_{2}}  =  4 z_{5}, \ \ \ \
\dot{z_{3}}  =  2 (z_{4} z_{7}-z_{5} z_{6}), \ \ \ \ 
\dot{z_{4}}  =  4 z_{2} z_{5}-z_{7}, \nonu \\
&&\dot{z_{5}}  =  z_{6}-4 z_{2} z_{4}, \ \ \ \  
\dot{z_{6}}  =  -z_{1} z_{5}+2 z_{2} z_{7}, \ \ \ \
\dot{z_{7}} =  z_{1} z_{4}-2 z_{2} z_{6}-4 z_{3}.
\label{eq:diff}
\eea
There are following five constants of motion of the system:
\bea
&&6 a  =  z_{1}+4 z_{2}^2-8 z_{4}, \ \
2 b  =  z_{1} z_{2}+4 z_{6}, \ \
c    =  z_{4}^2+z_{5}^2+z_{3}, \nonu \\
&&d    =  z_{4} z_{6}+ z_{5} z_{7}+z_{2} z_{3}, \ \
e    =  z_{6}^2+z_{7}^2- z_{1} z_{3}.
\label{eq:integral}
\eea
Although the Lax operator for this system is not readily available, 
we can still apply the asymptotic method due to Kowalewski to this system.
The integrable system generally possesses the Painlev\'e property,
i.e. solutions have only movable poles in the complex plane. 
We thus take $z_i=t^{-n_i} \sum_{j=0}^{\infty} A^{i}_{j} t^{j}$ where
$n_i$'s are positive integers\cite{SE,BvM}. 
Substituting these Laurent expansions
into the system of Eqs.(\ref{eq:diff}) and (\ref{eq:integral}),
one finds $n_i=1$ for $i=1, 2, 3$,
$n_i=2$ for $i=4, 5, 6, 7$ and a relation between the coefficients of 
$A^{i}_{j}$'s.
Then we obtain
the Laurent solutions for this system with seven parameters, 
five of which are from the constants of motion, $a, b, c, d, e$
and two additional ones $x$ and $y$ where they satisfy the equation
for an hyperelliptic curve\cite{BvM}:
\bea
y^2=(2 x^3-3 a x+b)^2 -4 ( 4 c x^2+4 d x+ e).
\label{eq:curve2}
\eea
We clearly see that with the following  substitution this gives the algebraic
curves given in Eq.(\ref{eq:curve}) of $N=2$ SUSY  $SU(3)$  
gauge theories with massive quarks of two flavors of masses $m_1$ and $m_2$:
\bea
&&y \rightarrow 2 y, \;\;\;\; a \rightarrow \frac{2}{3} u_{2}, 
\;\;\;\;
b \rightarrow -2 u_{3}, \ \ \ \
c \rightarrow \frac{1}{4} \Lambda_{2}^{4}, 
\nonu \\
&&d \rightarrow
\frac{\Lambda_{2}^{4}}{4} (m_{1}+m_{2}), \;\;\;\;
e \rightarrow \Lambda_{2}^{4} m_{1} m_{2}.
\label{eq:eq19}
\eea
When we consider the case of $c=0$, then this
leads to gauge theory coupled to one massive quark of 
mass $m_1$ or massless one$(N_f=1)$
after similar substitution. 
For the case of $c=d=0$,
the usual periodic Toda lattice
is recovered, and for $d=e=0$ we get back GC top.
So clearly we have a unifying model of two seemingly different systems.

\section{String Theory}
\noindent

Now let us consider the point like limit of string theories, where
the $N=2$ SUSY QCD is embedded in a compactification of string theory.
We obtain exact nonperturbative point particle limit  of a four
dimensional $N=2$ SUSY compactification of heterotic strings.
Using Heterotic/type II duality, 
we show how $N=2$ SUSY QCD 
with one flavor of massless quark arises
in type II string compactification on Calabi-Yau manifolds\cite{COFKM,HKTY,CFKM}.

Such analyses were performed for the following two cases\cite{KV}:
First is the case where the $E_8\times E_8$ heterotic string compactified
on $K3\times T^2$ is dual to the type IIB(or type IIA) 
theory compactified on a CY manifold (or its mirror), which is 
the weighted projective space of weights 1,1,2,2,6. 
The point like limit of this model was shown to yield the exact results
of Seiberg and Witten with pure $N=2$ Yang-Mills theory with gauge group $SU(2)$.
Second  case is  where the weighted projective space has weights 
1,1,2,8,12,  the point like limit is known to be the that of pure 
$N=2$ $SU(3)$ Yang-Mills theory.
By going to the conifold locus of the CY manifold and blowing it
up,  one can indeed obtain the algebraic curves for all the cases of  
$SU(n)$ gauge groups\cite{KLMVW}.

In order to relate these gauge theories with {\it matter} 
with string compactification scheme on a CY manifold, 
we look for the known cases where the
explicit forms of the discriminant and the Picard-Fuchs operators of the
CY manifolds have been worked out. 
One of the strong candidate is that of the weighted projective space with
weights 1,1,1,6,9.
This is because if we look at the discriminant locus in term of the coordinates
describing the large moduli parameters, 
the singularity structure of this is identical to that of 
$N=2$ $SU(2)$ gauge theory coupled to single $(N_f=1)$ flavor 
in the fundamental representation\cite{KV}. 

Consider the moduli space of the mirror of the weighted projective
space with weights 1,1,1,6,9 CY manifold with Hodge numbers 
$h_{1,1}=2, \;\; h_{2,1}=272$ whose defining polynomial given 
as follows\cite{CFKM}:
\beq
p = x_1^{18} + x_2^{18} + x_3^{18}+x_4^{3}+x_5^{2}
-18\ps x_1 x_2 x_3 x_4 x_5 - 3\ph x_1^6 x_2^6 x_3^6=0.
\eeq
This CY manifold has 2 vector multiplets whose scalar 
expectation values correspond
to $\ps$ and $\ph$ and 273 hypermultiplets including a dilaton field.
It is convenient to introduce the following variables that were used for
the complex moduli of the mirror:
\beq
x = \frac{3\ph}{(18\ps)^6},\ \ \ y = \frac{1}{(3\ph)^3}.
\eeq
The discriminant can be written as\cite{CFKM,HKTY}:
$
\Delta = (1-\bar{x})^3 - \bar{x}^3 \bar{y},
$
where $\bar{x} =2^4 3^3 x , \;\; \bar{y}=3^3 y$.
For weak coupling, $\bar{y} \rightarrow 0$, there exists a triple singularity at $\bar{x}=1$.
The locus on which the CY manifold aquires a conifold point is where $\De = 0$.

In order to go to the point like limit of strings ($\al ' \rightarrow 0$)
we would like to identify $\bar{x} -1$ with the vacuum expectation value of 4D gauge 
theory $u$ upto leading order of  $\alpha'$. In fact, to be dimensionally correct we need
\beq 
\bar{x} = 1 + \al' u + {\cal{O}}(\al'^2)=1+\frac{\ep}{\La_1^2} u +{\cal{O}}(\ep^2),
\eeq
where $\La_1$ is the renormalization scale parameter of the
theory with $N_f=1$.
At the conifold locus, we have
\beq
\bar{y} = \frac{(1-\bar{x})^3}{\bar{x}^3}\frac{1}{u^3}.
\eeq
When we expand for $\ps$ and $\ph$ we get
\beq
\ps = \frac{1}{18} \ep^{-\frac{1}{6}}( 1+ \ep \ps_1 + \cdots) , 
\ \ \  \ph = \frac{1}{3} \ep^{-1} (1+ \ep u+ \cdots),
\eeq
where $\ps_1 $ is independent of $u$.
With the expressions in the defining polynomial, we can now
compare with the the curve of SUSY QCD\cite{KHU}.
From the requirement that the coefficient of the term linear in $u$ be order of
$\ep$, 
we immediately see that the product of $x_1^6 x_2^6 x_3^6$ should be order of $\ep$.
Taking the following expansion, 
\bea
&&x_1  =  \ep^{\frac{1}{18}} a_1+\cdots, \;\;\; x_2=\ep^{\frac{1}{9}} a_2+\cdots ,\ \ \
x_3  = a_3(1+\ep b_3+\cdots), \nonu \\
&& x_4=a_4(1+\ep b_4+\cdots), \ \ \  
x_5 =  a_5(1+\ep b_5+\cdots),
\eea
and by requiring that $p$ has the following form up to the first power of 
$\ep$, we recover the hyperelliptic curve for $SU(2)$ $N_f=1$ gauge theory: 
\beq
p = \ep \left( 2u - 2x^2 + \hat{z} +\frac{\La_{1}^{3}(x+m)}{\hat{z}}+v^2+w^2\right) + 
{\cal{O}}(\ep^2) ,
\eeq
once we fix the functions in an appropriate form\cite{KHU}.
The change of variable $y=\hat{z}-P_2(x)$ gives rise to the explicit form
of the curve given in Eq.(\ref{eq:curve}).

Now we consider the periods.
As is the case of pure SUSY Yang-Mills theory\cite{KLMVW}, $p=0$ differs from 
(\ref{eq:curve}) by quadratic terms.
On the other hand,
the holomorphic 3-form\cite{CFKM} is
\beq
\Om = d \left( \ln \frac{\hat{z}}{\sqrt{Q(x)}
} \right) \wedge\left[\frac{dv\wedge dx}{\frac{\partial p}
{\partial w}}\right] .
\label{eq:3form}
\eeq
In order to integrate $\Om$ over $v$,  we solve for $w$ from $p=0$, and
plug this value of $w$ into (\ref{eq:3form}). Then the integral over $v$
becomes trivial, leading to the following result:
\beq
\int_y \Om =dx d \ln \frac{\hat{z}}{\sqrt{Q(x)}}=
d \left( x\;d \ln \frac{\hat{z}}{\sqrt{Q(x)}} \right)
\eeq
Now we see that the integral of $\Om$ on a 3-cycle of 
the CY manifold produces an integral of $d S$ over the cycle of Riemann surface.

\section{Conclusion}
\noindent

If one wishes to obtain the
prepotentials which are needed for exact effective action in 
SUSY gauge theory, we should consider quasiclassical $\tau$
fuctions in the context of integrable theory as in the case of pure
gauge theory\cite{NT}.
Of course the Nahm's equation can be subjected to averaging.
There are algebraic curves for higher rank cases with generic $N_c$ and 
$N_f$. Does this mean that a `higher' dimensional  generalization of GC top exists?
Although there exists multi-dimensional generalization\cite{BRS} of Kowalewski 
top, it is not available for GC top yet.
The relation to the Nahm's equation might be helpful here.
String duality and integrability of SUSY gauge models are closely related, but
still    needs    further    systematic     investigation,    especially    when    we    have 
matter\cite{KHU,KLMVW}.

\nonumsection{Acknowledgements}
\noindent
Part of  this is based on the works done in collaboration with Changhyun Ahn 
and S. Hyun. I thank Piljin Yi who explained the moduli space of magnetic monopoles.
I would like to thank A. Chodos for hospitality at Yale while this work was done.
This work is supported in part by Ministry of Education (BSRI-96-2442), 
KOSEF-JSPS exchange program, KOSEF 961-0201-001-2,  and by CTP/SNU.

\newpage
\nonumsection{References}

\end{document}
\end{thebibliography}
\end{document}